\documentclass[a4paper,fleqn,twoside]{article}

\usepackage{espcrc2mm}
\usepackage{graphicx}
\usepackage{amsmath}
\usepackage{amssymb}
\usepackage{isotope}

\title{Three-body vs.\ dineutron approach to two-neutron radiative capture in
$^{6}$He}

\author{%
L.V.~Grigorenko,\address[dub]{Flerov Laboratory of Nuclear Reactions, JINR,
141980
Dubna, Russia}$^,$%
\address[mifi]{National Research Nuclear University ``MEPhI'', Kashirskoye
shosse 31, 115409 Moscow, Russia}$^,$%
\address[kur]{National Research Centre ``Kurchatov Institute'', Kurchatov sq.\
1, 123182 Moscow, Russia}
N.B.~Shulgina,\addressmark[kur]$^,$%
\address[bog]{Bogoliubov Laboratory of Theoretical Physics, JINR, 141980 Dubna,
Russia}
M.V.~Zhukov\address[chal]{Department of Physics, Chalmers University of
Technology, S-41296 G\"{o}teborg, Sweden}
}

\begin{document}

\date{\today. {\tt .../latex/6he-sdm-2/6he-sdm-2-3.tex}}

\begin{abstract}

The low-energy behavior of the strength function for the $1^-$ soft dipole
excitation in $^{6}$He is studied theoretically. Use of very large basis sizes
and well-grounded extrapolation procedures allows to move to energies as small
as 1 keV, at which the low-energy asymptotic behavior of the E1 strength
function seems to be achieved. It is found that the low-energy behavior of the
strength function is well described in the effective three-body ``dynamical
dineutron model''. The astrophysical rate for the $\alpha$+$n$+$n \rightarrow
^6$He+$\gamma$ is calculated. Comparison with the previous calculations is
performed.

\vspace{1.5mm}


\vspace{1.5mm}

\noindent \textit{Date:} \today.

\end{abstract}

\maketitle

%===============================================================================

\section{Introduction}

%===============================================================================

The astrophysial radiative capture rates $\left \langle \sigma _{\text{capt}, 
\gamma } v \right \rangle$ are prime ingredients of the network nucleosynthesis 
calculations in the thermalized stellar envieronment. Some rates may be directly 
derived from experimental data. Some of them require sophisticated theoretical 
calculations and development of the adequate theoretical methods is essential in 
such cases.

\begin{figure}
\begin{center}
\includegraphics[width=0.45\textwidth]{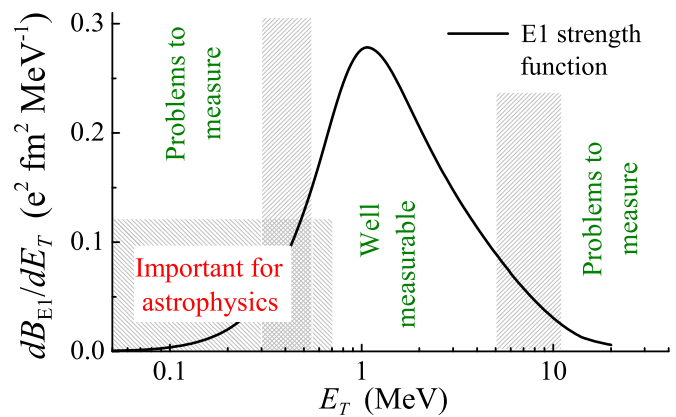}
\end{center}
\caption{(Color online) Schematic view of the soft dipole strength functions and
energy ranges available for measurements and important for astrophysics.}
\label{fig:scheme}
\end{figure}

The ability to reproduce in one theoretical calculation the behavior of the
electromagnetic strength function \emph{simultaneously} at intermediate energies
$E_T \sim 0.5-5$ MeV and at very low energies $E_T \lesssim 0.1-0.5$ MeV is
crucial for determination of the low-temperature astrophysical capture rates 
based on experimental data ($E_T$ is energy relative to the corresponding 
breakup threshold). The common idea is to measure the electromagnetic cross 
section at reasonably high energy (where it is relatively high) and then to 
extrapolate it to low energy theoretically, see Fig.\ \ref{fig:scheme}. For 
two-body radiative captures $A_1+A_2 \rightarrow A_{12} + \gamma$ this 
extrapolation is quite straightforward, which can be illustrated by analytical 
R-matrix type expression
\begin{equation}
\frac{d \sigma_{A_1A_2,\gamma}}{d E_T} \sim \frac{\Gamma(E_T)}{(E_T-E_r)^2 +
\Gamma^2_{\text{tot}}/4}\, , \qquad  \Gamma(E_T) \sim P_l(E_T)\, ,
\label{eq:2-b-extrapol}
\end{equation}
where the low energy asymptotic behavior is defined by the penetrability
function $P_l$ with definite angular momentum $l$. Obviously, this expression is
valid for resonant radiative capture. For nonresonant captures the direct
calculation of the electromagnetic strength function (SF) $dB_{\pi
\lambda}/dE_T$ of relevant multipolarity $\pi \lambda$ is required. However,
qualitative (especially, the low-energy) behavior of this SF is still mainly
determined by the penetrability function $P_l$.

%===============================================================================
%===============================================================================

For the three-body radiative captures the situation is far not that
straightforward. Since the classical paper \cite{Fowler:1967} and till the
modern compilation \cite{Angulo:1999} the semiclassical expression for
\emph{two-step} capture is commonly used for determination of the three-body
rates $A_1+A_2+A_3 \rightarrow A_{123} + \gamma$,
\begin{equation}
\left \langle \sigma _{A_1A_2A_3,\gamma }v \right \rangle = \sum_i \frac{\left
\langle \sigma _{A_1A_2,(A_1A_2)}v \right \rangle_i}{\Gamma_{(A_1A_2),i}}\,
\left \langle \sigma _{(A_1A_2)A_3,\gamma }v \right \rangle_i \,, \,
\label{eq:3b-rate-1}
\end{equation}
where $i$ is the number of the intermediate resonance populated at the first
step of capture into $(A_1 A_2)$ subsystem. This expression is obtained from the
rate equations for balance of three particles $(A_1A_2 A_3)$
\begin{eqnarray}
\dot{Y}_{(A_1A_2)}^{(i)}  & =& N_{A} \, \rho \;\left\langle
\sigma_{A_1A_2,(A_1A_2)}v \right \rangle
_{i}\;Y_{A_1}Y_{A_2} \nonumber  \\
& - & \Gamma_{(A_1A_2),i} \, Y_{(A_1A_2)}^{(i)} \,, \nonumber  \\
\dot{Y}_{(A_1A_2A_3)}  & =&\sum_{i} N_{A}\, \rho \;\left \langle
\sigma_{(A_1A_2)A_2,\gamma} v \right \rangle _{i}\;Y_{(A_1A_2)}^{(i)}Y_{A_3}\,,
\qquad
\label{eq:sys-eq}
\end{eqnarray}
where $Y^{(i)}_{A}$ are abundancies of the species $A$ in the state $i$, $\rho$
is density of the stellar media and $N_A$ is Avogadro constant. Equation
(\ref{eq:3b-rate-1}) arises under the assumption of thermodynamic equilibrium
for the intermediate resonant states: $\dot{Y}_{(A_1A_2)}^{(i)} \equiv 0$. Thus,
the ratio
\[
\left \langle \sigma _{A_1A_2,(A_1A_2)} \, v \right 
\rangle_i/\Gamma_{(A_1A_2),i}
\]
determines the classical concentration of the subsystem $(A_1A_2)$ in the
resonant state number $i$ in stellar media. Being essentially classical, the
Eq.\ (\ref{eq:3b-rate-1}) does not hold for a number of genuine
quantum-mechanical situations. Example of such a situation is the direct $2p$
radiative capture, which is the reciprocal process of $2p$ radioactive decay
\cite{Grigorenko:2005a}.

To formally generalize Eq.\ (\ref{eq:3b-rate-1}) for nonresonant capture rates
\cite{Fowler:1967,Angulo:1999} it is implicitly assumed that the ratio
\begin{equation}
\frac{\sigma _{A_1A_2,(A_1A_2)}(E) \, v(E) }{\Gamma_{(A_1A_2)}(E)} \,,
\label{eq:3b-rate-2}
\end{equation}
can be interpreted as the classical concentration of composite subsystems
$A_1+A_2$ at any given energy $E$ smaller than any resonance energy in the
system. It was found that although this idea \emph{qualitatively} looks quite
reasonable, the direct three-particle calculations can reveal important
\emph{quantitative} effects
\cite{Grigorenko:2005a,Grigorenko:2006,Grigorenko:2007a,Parfenova:2018}.

As a rule, the prevailing contribution to three-body non-resonant capture in a
wide temperature  range gives the dipole transition E1. Thus, the  problem of
three-body rates is connected with studies of soft dipole excitations (or soft
dipole mode, SDM) in halo systems. In the papers
\cite{Grigorenko:2006,Grigorenko:2007a,Parfenova:2018} we focused on the $2p$
captures, studied by the example of the $^{15}$O+$p$+$p \, \rightarrow
\,^{17}$Ne+$\gamma$ reaction. It was found that semisequential dynamics
(governed by the lowest resonances in the core+$p$ subsystem) is essential for
the low-energy behavior of the E1 SF determining the rate for this reaction. In
this work we studied the $2n$ captures for the case of the
$\alpha$+$n$+$n \, \rightarrow \,^{6}$He+$\gamma$ reaction. We find that for
the $2n$ captures the situation is qualitatively different: the low-energy
behavior of the E1 SF here is governed by the dynamics of the virtual state
(spin-singlet $s$-wave scattering) in the $n$-$n$ channel.

The astrophysical site, where $\alpha$+$n$+$n \, \rightarrow \,^{6}$He+$\gamma$ 
reaction (and analogous two-neutron captures) may become important is the 
r-process of nucleosynthesis in neutron-rich stellar media in conditions of high 
density, which makes possible three-body radiative captures. At the same time, 
the temperature should not be too high to avoid the inverse process of the 
photodisintegration. Several scenarios were suggested by astrophysicists: (i) 
the neutrino-heated hot bubble between the nascent neutron star and the 
overlying stellar mantle of a type-II supernova, (ii) the shock ejection of 
neutronized material via supernovae, (iii) merging neutron stars. Environment 
conditions such as temperatures and densities for these scenarios are quite 
different. For details see Ref.\ \cite[and Refs.\ 
therein]{Efros:1996,Bartlett:2006}. Calculations for specific scenarios may be 
the subject of separate studies.

There is a big difference in theoretical estimates of the $2n$ capture rates
for the $\alpha$+$n$+$n \, \rightarrow \, ^6$He+$\gamma$ reaction: the results
of papers
\cite{Efros:1996,Bartlett:2006,Goerres:1995a,deDiego:2010,deDiego:2011,%
deDiego:2014} are highly inconsistent with each other. Important motivation of
this work is also to get out of this uncertain situation.

%===============================================================================
\begin{figure}
\begin{center}
\includegraphics[width=0.49\textwidth]{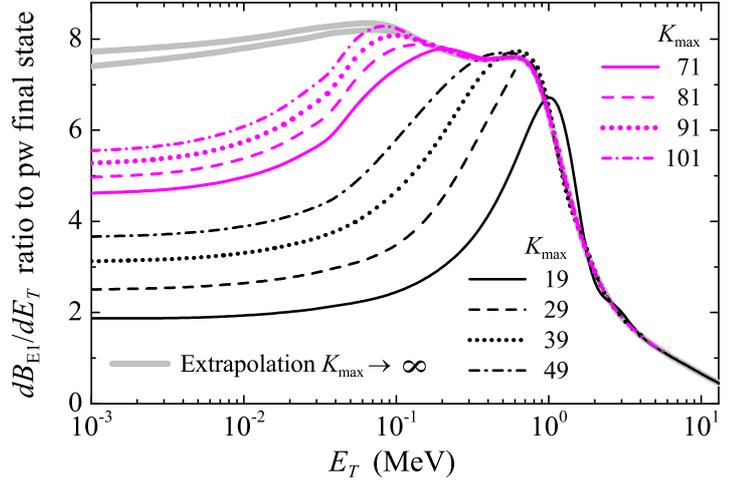}
\end{center}
\caption{(Color online) Low-energy ratio of the E1 SF calculated with full 
three-body Hamiltonian to that obtained in the ``no FSI'' approximation (plane 
wave final state is used). Curves correspond to different sizes $K_{\max}$ of 
the hyperspherical basis. Gray curves correspond to exponential extrapolation to 
infinite basis, see Fig.\ \ref{fig:basis-converg} (upper and lower boundaries, 
defined by the extrapolation uncertainty). See also Fig.\ 4 of
Ref.\ \cite{Grigorenko:2020}.}
\label{fig:rel-to-pw}
\end{figure}
%===============================================================================

%===============================================================================

\section{Low-energy convergence of the E1 SF}

%===============================================================================

The soft dipole excitation in $^{6}$He was studied in details in the recent
paper \cite{Grigorenko:2020}. For studies of E1 excitation the inhomogeneous
three-body Schr\"odinger equation is solved
\begin{eqnarray}
\left[ \hat{H}_3 + \tilde{V}_3(\rho)-E_T \right]  \Psi_{M_i m}^{JM(+)} =
\mathcal{O}_{\text{E1},m} \Psi^{J_{i}M_i}_{\mbox{\scriptsize gs}} \,, \quad
\nonumber \\
\hat{H}_3 = \hat{T}_3 + V_{cn_1}(\mathbf{r}_{cn_1}) +
V_{cn_2}(\mathbf{r}_{cn_2}) + V_{n_1n_2}(\mathbf{r}_{n_1n_2})\,.\quad
\label{eq:shred-e1}
\end{eqnarray}
providing the WF $\Psi_{M_i m}^{JM(+)}$ with pure outgoing wave asymptotics. The 
E1 transition operator has the following form
\[
\mathcal{O}_{\text{E1},m}=e \, \sum_{i=1,3}
Z_{i}\, r_{i} \,Y_{1 m}(\hat{r}_{i}) \,,
\]
and $\Psi^{J_{i}M_i}_{\mbox{\scriptsize gs}}$ is the $^{6}$He g.s.\ WF. The 
three-body potential $\tilde{V}_3$ provides phenomenological way to take into 
accound the many-body effects in three-cluster system, which are beyond the 
three-cluster approximation. The possible effect of this potential was shown to 
be not very important in \cite{Grigorenko:2020} and we neglect it in this work 
as well. The E1 strength function is then expressed via outgoing flux $j$ 
associated with the WF $\Psi_{M_i m}^{JM(+)}$:
\begin{equation}
\frac{dB_{\text{E1}}}{dE_T} = \frac{1}{2 \pi} \sum_J \frac{2J+1}{2J_i+1} \, j_J
\, .
\label{eq:dbde-1}
\end{equation}
The hyperspherical expansion of the continuum WF
\begin{equation}
\Psi_{ M_{i}m}^{JM(+)}=C_{J_{i}M_{i}1m}^{JM} \, \rho^{-5/2} \sum_{K \gamma}
\chi_{JK \gamma}^{(+)}(\varkappa \rho) \, \mathcal{J}_{K \gamma
}^{JM}(\Omega_{\rho}) \,,\quad
\label{eq:wf-cont}
\end{equation}
is truncated in our calculations by the maximum value of the generalized angular 
momentum $K=K_{\text{FR}}$. However, also the effective three-body potentials 
are used when solving Eq.\ (\ref{eq:shred-e1}), which are obtained by adiabatic 
procedure (so called ``Feshbach reduction'') and this procedure allowes to use 
much larger effective basis sizes $K=K_{\max}$.

%===============================================================================
\begin{figure}
\begin{center}
\includegraphics[width=0.49\textwidth]{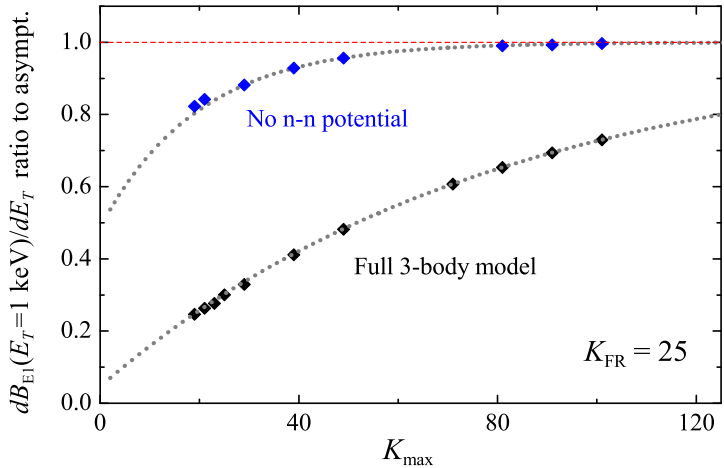}
\end{center}
\caption{(Color online) Example of the convergence of the E1 SF for $E_T=1$ keV
calculated in full three-body model and with $n$-$n$ FSI switched off 
(diamonds). Dotted curves show exponential extrapolation to infinite basis by 
Eq.\ (\ref{eq:exp}).}
\label{fig:basis-converg}
\end{figure}
%===============================================================================

It was shown in \cite{Grigorenko:2020} that the increasingly large size of
hyperspherical basis is needed to obtain converged E1 SF when moving to
lower energies, see Figs.\ 3 and 4 of \cite{Grigorenko:2020}. Visually converged
E1 SF was obtained in the whole energy range. However, if we investigate the
extreme low-energy part of the SF (also the range, important for astrophysical
calculations) we can find that the problem persists. One may see in Fig.\
\ref{fig:rel-to-pw} that even in the largest-basis calculations of 
\cite{Grigorenko:2020} with $K_{\max}=101$ the SF is converged down to $E_T \sim 
60-80$ keV. At lower energies (e.g.\ at $E_T=1$ keV), the curves corresponding 
to $K_{\max}=101$, 91, 81 are nearly equidistant indicating very slow 
convergence at maximum $K_{\max}$ achieved in the calculations.

What to do in this situation? The practical solution which we have already used
in the studies of the poorly converged two-proton widths (see, e.g. Refs.\
\cite{Grigorenko:2007,Brown:2014,Brown:2015,Grigorenko:2017}) is to use the
convergence trend for hyperspherical basis. It can be seen in Fig.\
\ref{fig:basis-converg} that the convergence over $K_{\max}$ has pefectly
exponential character
\begin{equation}
\frac{dB_{E_1}(E_T,K_{\max})}{dE_T} = \frac{dB_{E_1}(E_T,\infty)}{dE_T} - c_1
\exp \left(-\frac{K_{\max}}{c_2}\right) \,,
\label{eq:exp}
\end{equation}
in a broad range of $K_{\max}$ values from about 35 to 101. The convergence
character  shows that enormous basis sizes are needed for complete
convergence at low $E_T$ values: at $E_T=1$ keV the $95\%$ convergence would be
achieved at $K_{\max} \sim 250$. Direct calculation is thus not an option in
such situation.

\begin{figure}
\begin{center}
\includegraphics[width=0.49\textwidth]{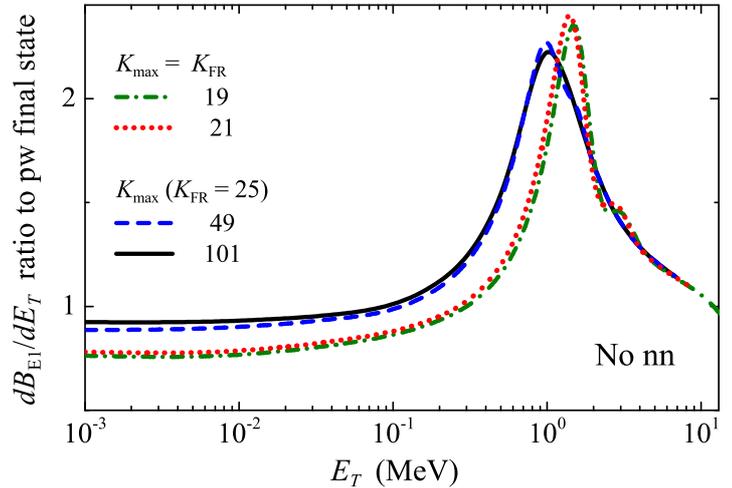}
\end{center}
\caption{(Color online) The same as Fig.\ \ref{fig:rel-to-pw}, but for ``no
$n$-$n$ FSI'' three-body Hamiltonian. See also Fig.\ 4 of Ref.\
\cite{Grigorenko:2020}.}
\label{fig:rel-to-pw-nonn}
\end{figure}

Where is the source of the convergence problem? We have found in
\cite{Grigorenko:2020} that the low energy convergence of the SF is much faster
if the $n$-$n$ interaction is switched off. The same calculations performed for
such a ``truncated'' Hamiltonian in the low-energy domain indicate that the
convergence issue is not severe in this case, see Fig.\
\ref{fig:rel-to-pw-nonn}. The calculations with the ``no $n$-$n$ FSI''
three-body Hamiltonian are fully converged (the $95\%$ convergence is achieved
with $K_{\max} \sim 45$, see Fig.\ \ref{eq:exp}). However, this approximation
provides drastically smaller ($\sim 9$ times) values of the E1 SF in the
low-energy domain, which shows that the $n$-$n$ FSI is essential for the
question.

%===============================================================================
%===============================================================================

%===============================================================================

\section{Dynamical dineutron model of SDM}

%===============================================================================

\begin{figure}
\begin{center}
\includegraphics[width=0.49\textwidth]{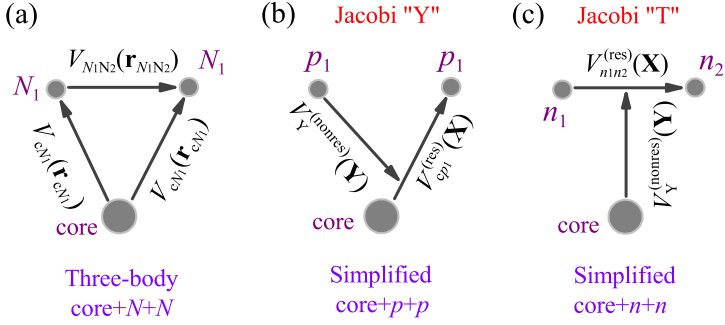}
\end{center}
\caption{(Color online) Simplification of the calculation scheme for SDM in the
three-body case. (a) Initial complete 3-body Hamoltonian. (b) For core+$p$+$p$
system the dynamical domination of resonances in the core-$p$ subsystem
motivates the use of simplified Hamiltonian in the ``Y'' Jacobi system. (c) For
core+$n$+$n$ system the dynamical domination of the $n$-$n$ FSI motivates the
use of simplified Hamiltonian in the ``T'' Jacobi system.}
\label{fig:three-schemes}
\end{figure}

Because the behavior of E1 SF in $^6$He is so sensitive to virtual
state in the spin-singlet $n$-$n$ channel, then maybe a good approximation to
it can be obtained by taking into account \emph{only} the dynamics of the
``dineutron''. This can be done applying the formalism developed in
\cite{Grigorenko:2006,Parfenova:2018} for studies of SDM excitations in
$^{17}$Ne, but in the ``T'' Jacobi system, see Fig.\ \ref{fig:three-schemes}.
What we get in this case can be called ``dynamic dineutron model''. Analogous 
model we have already applied for qualitative studies of two-neutron emission in 
dineutron approximation \cite{Grigorenko:2018}.

The idea of the method is that for E1 excitation studies instead of solving the
three-body Schr\"odinger equation Eq.\ (\ref{eq:shred-e1}) with Hamiltonian 
$\hat{H}_3$ we introduce the simplified Hamiltonian
\begin{equation}
\hat{H}'_3 = \hat{T}_3 + V_{y}(\mathbf{Y}) + V_{n_1n_2}(\mathbf{X})\,,
\label{eq:ham-simpl}
\end{equation}
which factorize the degrees of freedom in the ``T'' Jacobi system, see Fig.\
\ref{fig:three-schemes}. The latter Hamiltonian allows exact semianalytical
solution, since it has Green's function of a simple analytical form, which
(schematically) looks like
\begin{equation}
G^{(+)}_{E_T}(\mathbf{XY},\mathbf{X'Y'}) = \frac{1}{2 \pi i} \int dE_x
G^{(+)}_{E_x}(\mathbf{X},\mathbf{X'}) G^{(+)}_{E_T-E_x}(\mathbf{Y},\mathbf{Y'})
\,, \nonumber
\label{eq:gf-anal}
\end{equation}
where $G^{(+)}_{E_x}(\mathbf{X},\mathbf{X'})$ and
$G^{(+)}_{E_T-E_x}(\mathbf{Y},\mathbf{Y'})$ are ordinary two-body Green's
functions of the $X$ and $Y$ subsystems. This approach can be justified if
the interactions $V_{cn_1}(\mathbf{r}_{cn_1})$ and $V_{cn_2}(\mathbf{r}_{cn_2})$ 
in (\ref{eq:shred-e1}) are not of a prime importance for the system dynamics and 
can be replaced with one effective interaction $V_{y}(\mathbf{Y})$. It can be 
seen in Fig.\ \ref{fig:three-schemes} that both in two-proton case (b) and in 
two-neutron case (c) the dynamically important (in both cases resonant) 
interaction is associated with $X$ coordinate, while de-facto insignificant 
interactions are ``hidden'' in the effective interaction $V_y$ depending only on 
the $Y$ coordinate. For technical details of the three-body method and dineutron 
approximation, see Refs.\ \cite{Grigorenko:2020,Grigorenko:2018}.

%===============================================================================
%===============================================================================

\begin{figure}
\begin{center}
\includegraphics[width=0.49\textwidth]{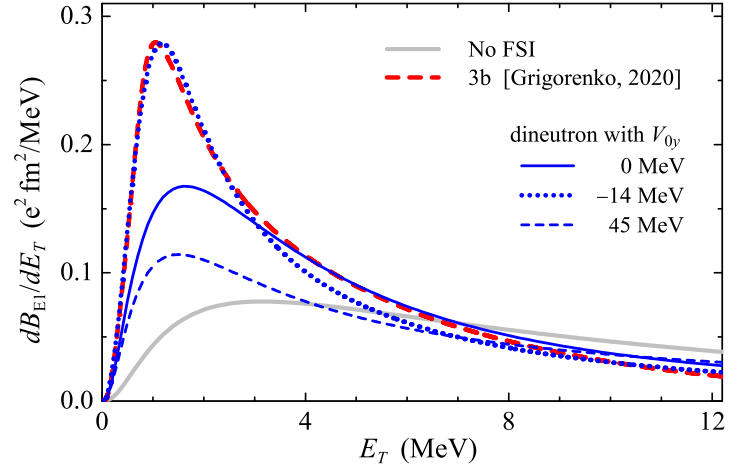}
\end{center}
\caption{(Color online) Comparison of the E1 strength functions calculated in
full three-body model (Grigorenko, 2020: \cite{Grigorenko:2020}), in ``no FSI'' 
approximation, and in different dineutron model settings.}
\label{fig:dipro-lin}
\end{figure}

\begin{figure}
\begin{center}
\includegraphics[width=0.49\textwidth]{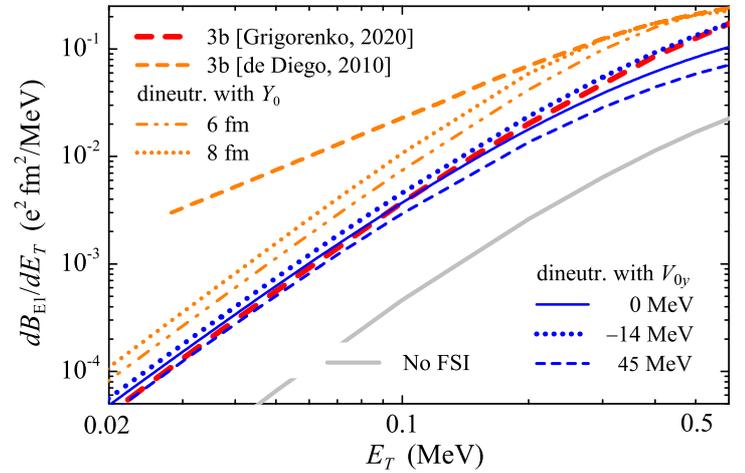}
\end{center}
\caption{(Color online) Comparison of the low-energy asymptotics of the E1
strength functions calculated in full three-body model (Grigorenko, 2020: 
\cite{Grigorenko:2020}), in ``No FSI''
approximation, in different dineutron model settings, and in paper (de Diego, 
2010: \cite{deDiego:2010}).}
\label{fig:dipro-log}
\end{figure}

The calculations of E1 SF within dynamical dineutron model are shown in Fig.\
\ref{fig:dipro-lin}. Three test interactions in the $Y$ subsystem have
the Gaussian formfactors
\[
V_{y}(Y) = V_{0y} \, \exp[-(Y/Y_0)^2]\,,
\]
with $Y_0=3$ fm, acting in $p$-wave only. They are: (i) no interaction
$V_{0y}=0$ (leads to plane wave over $Y$ coordinate), (ii) attraction with
$V_{0y}=-14$ MeV, and (iii) repulsion with $V_{0y}=45$ MeV. Attractive
interaction was fitted to reproduce the profile of the three-body E1 strength
function in a broad energy range. However, if we turn to low-energy behavior of
the E1 SF in Fig.\ \ref{fig:dipro-log}, then we see that the best match with
calculated low-energy behavior of a three-body SF is obtained with repulsive
$V_{y}$ potential. The ``trivial'' assumption of the absence of interaction
$V_{0y}=0$ in $Y$ subsystem leads to overall good agreement with the
three-body SF. In any case a comparison of attractive and strongly repulsive 
interactions shows a mismatch of only $\lesssim 50 \%$ in the low-energy region. 
Therefore, the uncertainty associated with the ``unphysical'' interaction $V_y$ 
is not large in the asymptotic region anyhow, although it changes drastically 
the profile of the E1 SF at higher energies.

%===============================================================================
%===============================================================================

The nearly linear behavior of the E1 SFs in the left part of log-scale Fig.\
\ref{fig:dipro-log} indicates that the correct low-energy asymptotic behavior
\begin{equation}
dB_{E_1}(E_T)/dE_T \sim E_T^3  \,,
\label{eq:pow}
\end{equation}
is almost achieved.

%===============================================================================
%===============================================================================

%===============================================================================

\section{Three-body capture rate}

%===============================================================================

The E1 nonresonant astrophysical radiative capture rate for the three-body
reactions is given by the expression
\begin{eqnarray}
\left \langle \sigma _{A_1A_2A_3,\gamma } v \right \rangle =  \left( \frac{\sum
A_i}{\prod A_i}\right)
^{3/2}\left( \frac{2\pi }{mkT}\right) ^{3}\;\frac{2(2J_f+1)}{\prod(2J_i+1)}
\nonumber \\
\times  \int dE_T \, \frac{16\pi}{9}\, E_{\gamma }^{3}\;
\frac{dB_{E1}(E_T)}{dE_T} \exp \left[ -\frac{E_T}{kT}\right] \,, \quad
\label{eq:nonres-rate}
\end{eqnarray}
where $E_{\gamma}=E_T+ E_b$ ($E_b=0.973$ MeV for $^6$He) and $J_i$ are the spins
of incident clusters, while $J_f$ is the spin of the bound final state ($0^+$ in 
the $^{6}$He case). Note that the E1 strength function $dB_{E1}/dE_T$ in Eq.\ 
(\ref{eq:nonres-rate}) is the strength function for the reciprocal process of 
$^{6}$He E1 EM dissociation.

The two-neutron capture rates calculated with SFs discussed above are shown in
Fig.\ \ref{fig:rate}. The most trivial dineutron model result with  $V_{0y} = 0$
has a good overall agreement with the three-body result (the deviation is never
more than $~50\%$). The temperature region from 1 to 10 GK is better described
by the dineutron model with $V_{0y} = -14$ MeV, reproducing best the ``bulk'' of
the three-body SF.

If we perform the rate calculations starting with the asymptotic expression for
the SF (\ref{eq:pow}) then the rate is given by
\begin{equation}
\left \langle \sigma _{2n,\gamma } v \right \rangle \sim T \left[
1+12(E_{\text{b}}/T) + 60(E_{\text{b}}/T)^2 + \ldots \right]   \,.
\label{eq:rate-pow}
\end{equation}
This asymptotic expression (shown by the green dashed curve in Fig.\
\ref{fig:rate}) is very precise up to $T \sim 0.1$ GK and at $T \sim 0.6$ GK the
difference from the three-body SF is just a factor of 2. This emphasizes the
importance of a correct description of the SF low-energy asymptotics.

Finalizing the discussion here, the phenomenological recipe for use of dineutron 
model seems very simple:

\noindent (i) If there is no experimental information at all, then it is very 
reasonable to make the rate estimates with $V_{y} \equiv 0$. As we have seen 
above, in the case of $^{6}$He the overall agreement in a broad temperature 
range is also very reasonable.

\noindent (ii) If there is experimental information about E1 strength function, 
parameters of the dineutroton model can be fitted to the experimental SF 
profile. In that case we have nearly perfect description of the rate at 
$T>0.7-1.0$ GK. At $T \rightarrow 0$ the dineutron model does not guarantee 
precise asymptotic behavior, but the mismatch is not severe.

\noindent (iii) Despite the  uncertainties of the dineutron model it immediately 
provides the results which is much closer to the highly accurate three-body 
calculation results than any result obtained in this field before, see Fig.\ 
\ref{fig:rate} and discussion of the next Section. In that sense it is very 
strong phenomenological tool.

%===============================================================================

\section{Comparison with previous results}

%===============================================================================

The calculations of the astrophysical radiative capture rate for the
$^{4}$He+$n$+$n\,\rightarrow \,^6$He+$\gamma$ reaction are given in a number of
papers \cite{Efros:1996,Bartlett:2006,Goerres:1995a,deDiego:2010,deDiego:2011,%
deDiego:2014}. The results of the papers
\cite{Efros:1996,Bartlett:2006,Goerres:1995a} are based on different
quasiclassical two-step approximations. So, maybe, it is not surprising that
they are highly incompatible with each other and with results of this work.

%===============================================================================
\begin{figure}
\begin{center}
\includegraphics[width=0.49\textwidth]{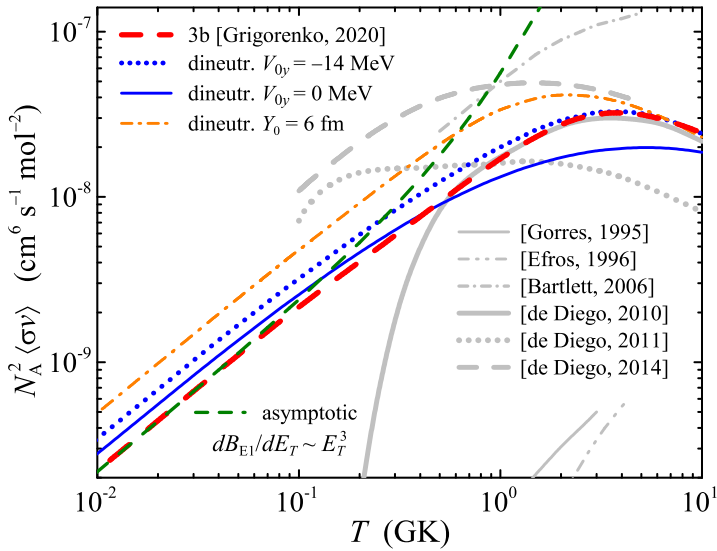}
\end{center}
\caption{(Color online) Three-body astrophysical radiative capture rates for the
$\alpha$+$n$+$n \rightarrow ^6$He+$\gamma$ reaction obtained with different E1
SFs in this work (colored curves) and in the other models (gray curves) in 
Refs.\ (Goerres, 1995: \cite{Goerres:1995a}), (Efros, 1996: \cite{Efros:1996}),
(Bartlett, 2006: \cite{Bartlett:2006}), (de Diego, 2010: \cite{deDiego:2010}), 
(de Diego, 2011: \cite{deDiego:2011}), (de Diego, 2014: \cite{deDiego:2014}).}
\label{fig:rate}
\end{figure}
%===============================================================================

More attention needs to be paid to the results of the three-body model
\cite{deDiego:2010,deDiego:2011,deDiego:2014}, which, in principle, should be
consistent with the results of this work. There are two issues.

\noindent (i) All three results \cite{deDiego:2010,deDiego:2011,deDiego:2014}
are declared to be based on the same E1 SF from paper \cite{deDiego:2010}.
However, different rate values can be found in papers
\cite{deDiego:2010,deDiego:2011,deDiego:2014}, see Fig.\ \ref{fig:rate}. We have
no understanding of this fact.

\begin{figure}
\begin{center}
\includegraphics[width=0.49\textwidth]{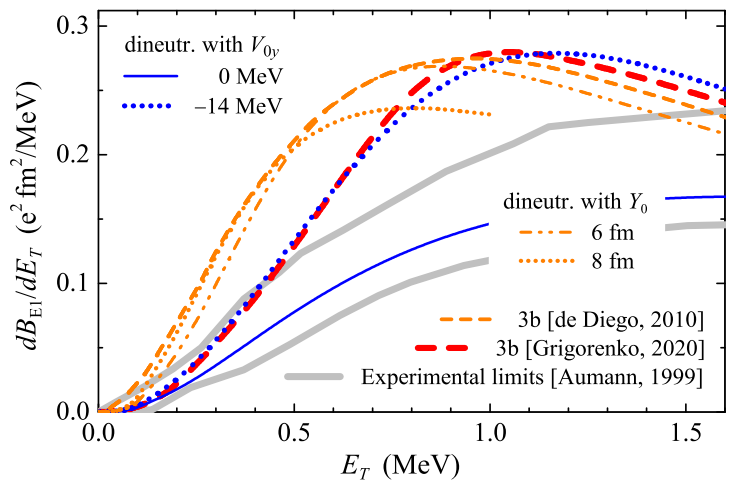}
\end{center}
\caption{(Color online) Comparison of the E1 low-energy strength functions
calculated in our full three-body model (Grigorenko, 2020: 
\cite{Grigorenko:2020}), in different dineutron model settings, in paper (de 
Diego, 2010: \cite{deDiego:2010}) with experimental data of Ref.\ (Aumann, 1999: 
\cite{Aumann:1999}).}
\label{fig:expcomp}
\end{figure}

\noindent (ii) It was discussed in Ref.\ \cite{Grigorenko:2020} that the E1 SF
from \cite{deDiego:2010} has some kind of suspicious enhancement of the
low-energy behavior, which is not reproduced in the other three-body approaches
see Fig.\ \ref{fig:expcomp} of this work, Fig.\ 14 in Ref.\ 
\cite{Grigorenko:2020}, and also Refs.\ \cite{Cobis:1997,Myo:2001}.

We attempted to reproduce the low-energy behavior of the E1 SF
\cite{deDiego:2010} in the dynamical dineutron model. That was found to be very
difficult. Evidently, the low-energy enhancement of the SF requires the
reduction of the centrifugal barrier in the $Y$ channel (for E1 transiiton the
``dineutron'' cluster should be in $l_y=1$ relatively $\alpha$-core). It can be
seen in Fig.\ \ref{fig:expcomp} that the SF, which is pretty close to
\cite{deDiego:2010}, can be obtained in the dineutron model. However, this
requires an unrealistic potential in $Y$ channel: here we use Gaussian potential
with extremely large radius of $Y_0 = 6$ fm, which in our opinion has no
reasonable justification.  And even so, if we look in Fig.\ \ref{fig:dipro-log}
it can be found that still it does not help to reproduce the correct asymptotic
low-energy behaviour of the E1 SF. Even more extreme potential, with $Y_0 = 8$
fm, is required to reproduce the behavior of SF from \cite{deDiego:2010} down to
$E_T \sim 0.3$ MeV and for lower energies the dineutron SF turns to expected
$\sim E^3_T$ trend. So, in the log-scale it can be seen that the low-energy SF
of \cite{deDiego:2010} has no chance to be reconciled with ours.

The rate calculated in \cite{deDiego:2010} overlaps with our three-body result
in a broad temperature range (and it is drastically smaller for $T<0.5$ GK). We 
think that this contradicts SF behavior. The dineutron SF with $Y_0
= 6$ fm approximates SF \cite{deDiego:2010} well: it is \emph{smaller or equal}
to SF \cite{deDiego:2010} in the whole energy range, see Fig.\
\ref{fig:expcomp}. However, the rate computed with this dineutron SF is
\emph{larger} than the rate from \cite{deDiego:2010} in the whole temperature
range, see Fig.\ \ref{fig:rate}.

%===============================================================================

\section{Conclusion}

%===============================================================================

The convergence of the SDM (E1) strength function for $^{6}$He becomes slower
with decreasing decay energies. Large-basis (with $K_{\max}=101$) calculations
allowed to obtain fully converged SF values down to energies as low as 60-80
keV. For the lower energies (e.g., as small as 1 keV) it was shown that the
extrapolation scheme allows to obtain reliable SF values.

%===============================================================================
%===============================================================================

It was demonstrated that the low-energy E1 SF in $^{6}$He case is strongly
affected by the virtual state in the spin-singlet $n$-$n$ channel. For that
reason a very reliable approximation for the low-energy E1 SF can be obtained in
a dynamical dineutron model. Within the dineutron approximation the three-body
dynamics is reduced to a kind of factorized two-body semisequential dynamics.
As a result, the three-body Green's function in the dineutron approximation
has a compact analytical form, allowing exact semi-analytical calculations. This 
is an important result in several ways:

\noindent (i) The dineutron model provide a simple seminalytical cross
check and reliable shortcut for the bulky three-body calculations for the
low-energy three-body (namely, two-neutron) radiative capture reactions.

\noindent (ii) Important \textit{qualitative difference} between two-proton and 
two-neutron radiative captures is elucidated, see Fig.\ \ref{fig:three-schemes}. 
In the case of the low-energy two-proton capture the dynamics is also factorized 
to two-body semisequential dynamics, but in the ``Y'' Jacobi system, which 
allows to take into account the low-lying
resonances in the core+$p$ channel. The diproton correlation does not
play important role in the low-energy region.

\noindent (iii) The effective low-energy
reduction of the three-body dynamics to dynamics of dineutron emission may be
seen as very intuitive and even trivial result. However, without bulky
three-body calculations we would have never be confident to which level of
precision this approach really works. Now, the semi-analytical  dineutron
model, supported by our high-precision three-body calculations, reliably
predicts the low-energy behavior of the strength function and capture rates and,
thus, provides reliable extrapolation of experimental data measured at
sufficiently high energies.

All the previous results
\cite{Efros:1996,Bartlett:2006,Goerres:1995a,deDiego:2010,deDiego:2011,%
deDiego:2014} for the $^{4}$He+$n$+$n\,\rightarrow \,^6$He+$\gamma$
astrophysical radiative capture rate are highly inconsistent with each other and
with the results of this work. For calculations
\cite{deDiego:2010,deDiego:2011,deDiego:2014} the origin of important problems
can be identified as inconsistent treatment of the low-energy region of the E1
SF. Thus, our results emphasize the importance of the accurate treatment of
few-body dynamics for consistent determination of the low-temperature parts of
the astrophysical three-body capture rates.

%===============================================================================

\paragraph*{Acknowledgments}

LVG was supported in part by the Russian Science Foundation grant No.\
17-12-01367.

%===============================================================================

%\bibliographystyle{apsrev4-1}
%\bibliography{d:/latex/all}

\bibliographystyle{elsart-num-m}
\bibliography{d:/latex/all}

%###############################################################################

\end{document}